\documentclass[aps,twocolumn,pr,superscriptaddress,longbibliography]{revtex4-1}
\usepackage{times,amsmath,graphicx}
\usepackage[colorlinks=true,citecolor=blue,urlcolor=blue,anchorcolor=blue,linkcolor=blue]{hyperref}
\usepackage{amssymb}
\usepackage{multirow}
\usepackage{xcolor}

\begin{document}
\title{New-record-$T_{c}$ and three-gap 2D superconductors with electronic and phononic topology: KB$_{2}$C$_{2}$}

\author{Hao-Dong Liu}
\affiliation{Institute of Applied Physics and Computational Mathematics and National Key Laboratory of Computational Physics, Beijing 100088, China}
\affiliation{Graduate School, China Academy of Engineering Physics, Beijing 100088, China}


\author{Xin-Peng Fu}
\affiliation{Institute of Applied Physics and Computational Mathematics and National Key Laboratory of Computational Physics, Beijing 100088, China}
\affiliation{Graduate School, China Academy of Engineering Physics, Beijing 100088, China}


\author{Zhen-Guo Fu}
\thanks{E-mail: fu\_zhenguo@iapcm.ac.cn}
\affiliation{Institute of Applied Physics and Computational Mathematics and National Key Laboratory of Computational Physics, Beijing 100088, China}
\affiliation{Graduate School, China Academy of Engineering Physics, Beijing 100088, China}

\author{Hong-Yan Lu}
\thanks{E-mail: hylu@qfnu.edu.cn}
\affiliation{School of Physics and Physical Engineering, Qufu Normal University, Qufu 273165, China}

\author{Ping Zhang}
\thanks{E-mail: zhang\_ping@iapcm.ac.cn}
\affiliation{Institute of Applied Physics and Computational Mathematics and National Key Laboratory of Computational Physics, Beijing 100088, China}
\affiliation{School of Physics and Physical Engineering, Qufu Normal University, Qufu 273165, China}

\date{\today}
\begin{abstract}
Pursuing higher-temperature superconductors under ambient pressure continues to be a prominent topic in materials discovery. Isomorphic structures like MgB$_{2}$ exhibit potential for conventional BCS-type superconductivity, but their transition temperatures ($T_{c}$) have remained below 100 K based on both experimental findings and theoretical predictions. In this study, two new two-dimensional (2D) superconductors with sandwich structures, KB$_{2}$C$_{2}$, featuring BC layers in AA and AB stacking configurations, are designed, whose $T_{c}$ can exceed 112 K, setting a new record in 2D superconductors. The analyses suggest that electrons in $\sigma$-states covalent bonds and high-frequency $E$ phonon modes dominated by the in-plane vibrations of B/C atoms are predominately responsible for electron-phonon coupling (EPC). An exciting robust three-gap superconducting nature stems from the strong and evident three-region distribution characteristic of electronic EPC parameters $\lambda^{\mathit{el}}_{\mathbf{k}}$. When biaxial tensile strain (BTS) is applied, their $T_{c}$ are boosted above 153 K. The increase in $T_{c}$ originates from the softening of optical $E$ phonon modes around the $\Gamma$ point and acoustic modes around the Q point, rather than an increase of electrons at Fermi level ($E_{F}$) as observed in other similar systems. Thus, phonon plays a more beneficial role in the EPC of BTS cases, highlighting its significance as a medium in BCS superconductors. Moreover, we find KB$_{2}$C$_{2}$ exhibits interesting topological properties, spin antivortex, and Ising-type spin splitting. This is the first report of the coexistence of nontrivial topology and superconductivity with such a high $T_{c}$. Therefore, KB$_{2}$C$_{2}$ may offer promising sandwich structures to explore higher-$T_{c}$ 2D superconductors, alongside present potential avenues for investigating fundamental quantum physics.

\end{abstract}
\maketitle

\begin{figure*}[tbph!]
	\centering
	\includegraphics[width=0.8\linewidth]{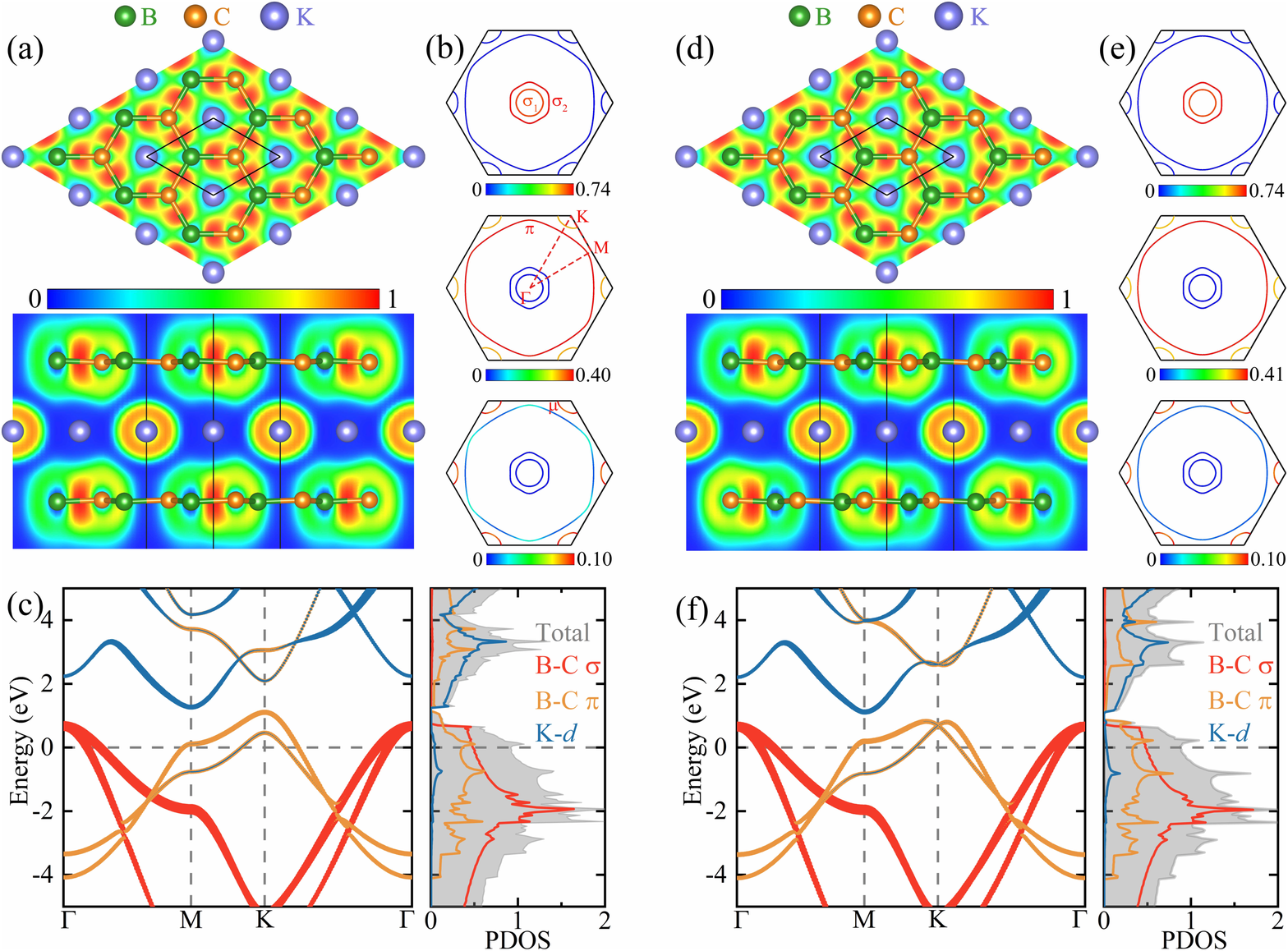}
	\caption{(a) Top and side views of AA-KB$_{2}$C$_{2}$ crystal structure with the electron localization function of B-C and (1 1 0) planes, respectively. The black solid rhombus is the unit cell, which contains two B, two C, and one K atoms. (b) Distributions of B-C $\sigma$, $\pi$, and K $d$ orbitals on the FS, namely $\sigma_{1}$, $\sigma_{2}$, $\pi$, and $\mu$, respectively. High symmetry routes and points are shown in red. (c) Electronic band structure and PDOS projected with orbitals. (d) $\sim$ (f) The corresponding to electronic properties of AB-KB$_{2}$C$_{2}$ with the color representing the contribution of each orbitals.}
	\label{fig:fig}
\end{figure*}
$Introduction.$ Ever since the discovery of superconductivity in 1911, a major objective for the fields of condensed matter physics, chemistry, and materials science has been to expand the varieties of high-$T_{c}$ superconductors. Over the past century, tremendous efforts have been focused on conventional superconductors based on the Bardeen-Cooper-Schrieffer (BCS) \cite{PhysRev.108.1175} theory. Particularly, in 2001, the discovery of two-gap \cite{Choi2002,PhysRevB.92.054516} superconductivity with 39 K in the fairly straightforward compound MgB$_{2}$ \cite{Nagamatsu2001,PhysRevLett.87.037001,PhysRevB.64.020501} further inspires enthusiasm for the exploration. Extensive researches have been conducted to understand the mechanism hiding in its superconductivity. It has been established that large electronic density of states (DOS) at the $E_{F}$ \cite{PhysRevB.64.020501}, significant phonon with high Debye frequency \cite{PhysRevLett.124.077001,PhysRevLett.86.4656}, and strong EPC between metallized $\sigma$ electrons coming from the $p_{x,y}$ orbitals \cite{PhysRevLett.86.4366,PhysRevLett.87.037001,PhysRevLett.86.5771,PhysRevLett.86.4656,PhysRevLett.87.087005} and the in-plane phononic vibrations in boron layers, especially the $E_{2g}$ \cite{PhysRevLett.87.087005} stretching modes, play critical roles. Moreover, the existence of multi-gap superconductivity formed from distinctly different orbital projections on the Fermi surfaces (FS) \cite{PhysRevLett.94.037004,PhysRevB.75.054508,PhysRevB.101.104507} is another crucial and indispensable approach for achieving high-$T_{c}$ superconductors. Based on these advantages, numerous new borides (Table S1) have been discovered through atomic doping and substitution methods. However, the $T_{c}$ of most borides are lower that the McMillan limit (39 K). 

In light of the fact, there has been a growing interest in the potential of graphite-like stacking borocarbides \cite{PhysRevB.89.045136,PhysRevB.101.094501,PhysRevB.102.144504,PhysRevB.104.054504,Singh2022,doi:10.1021/acs.nanolett.2c05038,PhysRevB.107.134502,PhysRevB.107.014508,PhysRevB.91.045132}, as a mean to discover new superconductors with higher $T_{c}$ at ambient pressure. The compounds not only meet the high-$T_{c}$ standards of BCS theory but also possess stronger covalent bonds originating from B and C atoms rather than only B atoms. The surfaceization of bulk phases is also a promising platform for the realization of high-$T_{c}$ 2D superconductors \cite{GINZBURG1964101,PhysRevB.7.3028,PLUMMER2003251,PhysRevLett.21.1320,Ueno2008,Ueno2011,PhysRevB.83.220503,PhysRevB.87.140503,10.1093/nsr/nwz204,PhysRevX.11.021065,Liu2022}. 

It is essential to consider whether any other materials could exhibit higher-$T_{c}$ superconductivity at ambient pressure. With that in mind, we carry out the systematical and comprehensive investigation of the superconductivity of borocarbides, specifically the sandwich KB$_{2}$C$_{2}$. The compounds consist of two honeycomb B-C layers intercalated by a hollow triangular potassium (K) layer. Using the first-principle calculations and the fully anisotropic Migdal-Eliashberg (ME) theory \cite{migdal1958interaction,eliashberg1960interactions,Philip,PhysRevB.87.024505}, we perform a through study for their crystal structures, basic and topological properties of electrons/phonons, and the nature of three-gap superconductivity. We find exciting results that the sandwich structures display three separated superconducting gaps with record-high $T_{c}$ \textgreater 112 K. Furthermore, the $T_{c}$ can be boosted to over 153 K under BTS.

\begin{figure*}[htb!]
	\centering
	\includegraphics[width=0.8\linewidth]{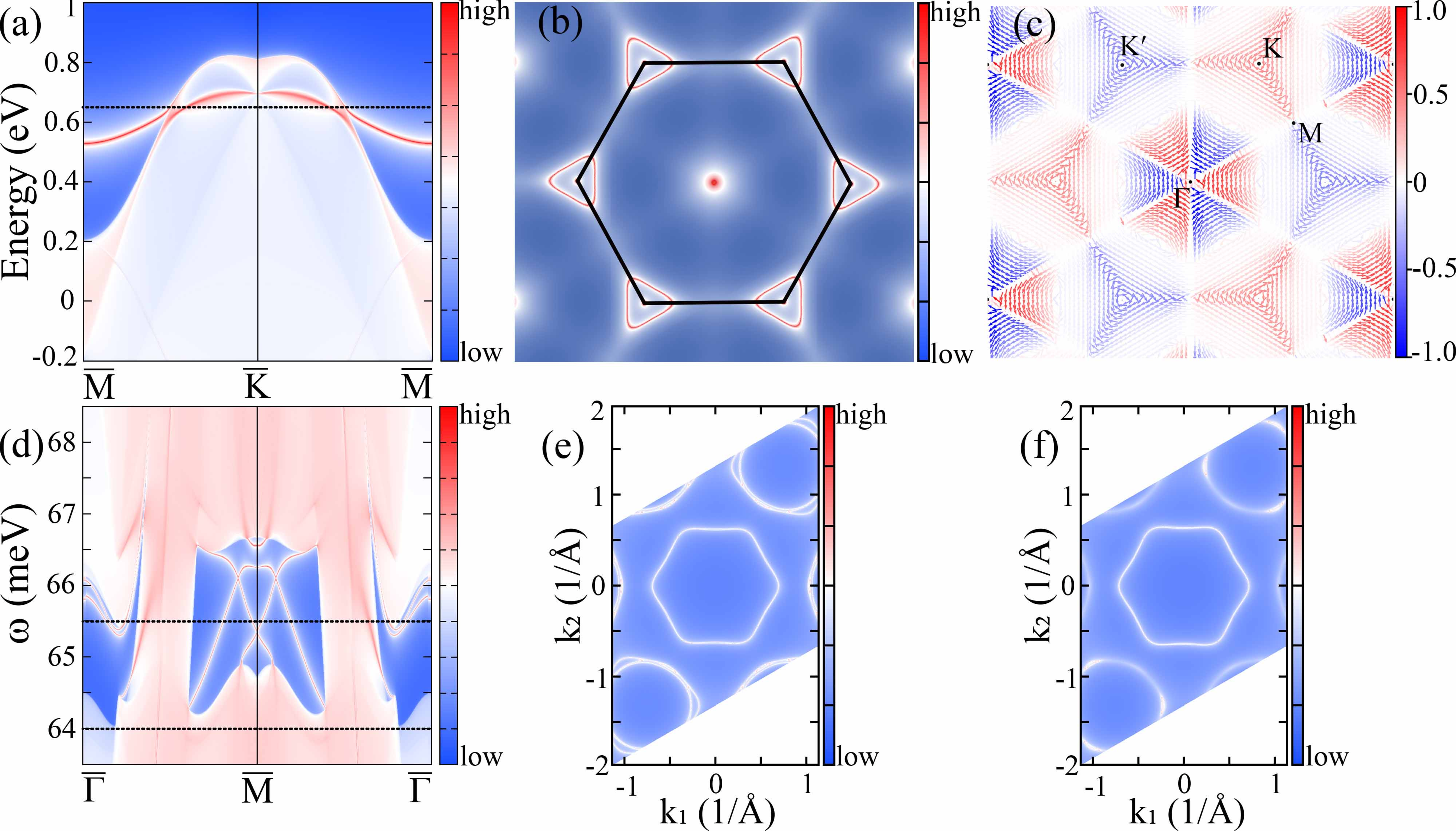}
	\caption{The topological behaviors of AB-KB$_{2}$C$_{2}$. (a) The electronic topological edge states on the (100) plane along the edge high-symmetry points $\bar{M}$-$\bar{K}$-$\bar{M}$. (b) The $k_{x}$-$k_{y}$ plane at energy levels equivalent to 0.65 eV in the whole BZ is marked with black dotted lines in (a). (c) The spin texture of the highest occupied band in the whole BZ under SOC. The arrows and the color depict in-plane spin and the $z$-component polarization, respectively. The zoom-in of spin texture around $K^{'}$, $K$, and $\Gamma$ points is shown in Fig. S6. (d) The phononic edge LDOS on the (100) plane between 63.5 and 68.5 meV along $\bar{\Gamma}$-$\bar{M}$-$\bar{\Gamma}$. (e) and (f) The corresponding isofrequency surface contours on (001) plane at 64 and 65.5 meV, respectively, marked with black dotted lines in (d).}
	\label{fig:fig}
\end{figure*}

We used the QUANTUM ESPRESSO (QE) \cite{Paolo} package and EPW code \cite{Giustino,Jesse,PhysRevB.87.024505} to calculate the phonon dynamical matrices and EPC, respectively. All the topological behaviors of AB-KB$_{2}$C$_{2}$ were performed with WannierTools \cite{WU2017,PhysRevLett.120.016401}. More computational methods and details are presented in Section II of Supplementary Materials (SM) \cite{SM}.

$The$ $stable$ $crystal$ $structures.$ The physical properties of materials are frequently determined by their dominant structural units. By precisely adjusting the geometries, one can dramatically modulate the electronic, phononic, and superconducting properties. Therefore, it is crucial to first comprehensively identify the dominant structural unit. 

Depicted in Fig. S1, we investigate the impact of several designs, including the face-to-face alignment, rotation, and gliding of B-C layers, on the construction of crystal structures. Among the investigated configurations, only the AA[Fig. S1(a)]- and AB[Fig. S1(d)]-KB$_{2}$C$_{2}$ possess dynamic stabilities. We also illustrate the time-dependent variations of the free energy (AIMD) for the two dynamic-stability configurations in Fig. S2, demonstrating their thermal stabilities at room temperature (300 K) and even at elevated temperatures up to 1300 and 1200 K, respectively. Our observations indicate that both configurations can maintain their sandwich structures at these elevated temperatures. However, exceeding the limits, as shown in Figs. S2(c) and S2(f), the K atoms can break free from the confinement of B-C layers, resulting in the disappearance of sandwich structures. Additionally, the formation/cohesive energy and mechanical stability can be obtained in Section III of SM \cite{SM}. To sum up, both investigated sandwich configurations of KB$_{2}$C$_{2}$ are stable and suitable for further study. 

The relaxed crystal structures of AA- and AB-KB$_{2}$C$_{2}$ are shown in Figs. 1(a) and 1(d), respectively. The B-C bond length is 1.60 \AA, smaller than the B-B bond length of 1.77 \AA \, in MgB$_{2}$. This can lead to a better orbital overlap between B and C atoms, subsequently strengthening the $\sigma$-state metallic covalent bonds that couple more effectively with phonon \cite{PhysRevB.107.134502}. More detailed crystal information such as space groups, Wyckoff position, lattice constants, and minor wrinkles between B and C atoms within the same layer, can be found in Table S2.

$Electronic$ $properties.$ Based on the Bader calculations \cite{HENKELMAN2006354} in Table S4, both sandwich structures of KB$_{2}$C$_{2}$ demonstrate the same quantitative charge transfer. In detail, one C atom gains 2.04 electrons, while each B/K atom loses 1.71/0.66 electrons. By analyzing electronic DOS [Figs. 1(c) and 1(f)] and charge density difference (Fig. S5), we further investigate the charge transfer: the B-C bonds and K atomic $d$ orbitals gain charges, while the centers of B-C graphene-like nets, the $p_{z}$ orbitals of the B/C atoms, and the $s$ orbitals of the K atoms lose charges. Subsequently, as shown in Figs. 1(a) and 1(d), the charges on the B-C bonds form strong covalent bonds, which are essential for high-$T_{c}$ superconductivity. Other orbitals with small contributions at $E_{F}$ are ignored, as depicted in Fig. S5. Lastly, the B-C $\sigma$ and $\pi$ states are dominant at $E_{F}$, with a minor contribution from K atomic $d$ orbitals. Due to lattice chemical pressure between B-C layers leading to an energetic level transition, part of K atomic valence electrons are transferred from $s$ to $d$ orbitals \cite{PhysRevB.104.054504,doi:10.1021/acs.nanolett.2c05038}. Four continuous bands cross the $E_{F}$ and show the equally quantitative pockets on the FS for both AA- and AB-KB$_{2}$C$_{2}$, as shown in Figs. 1(b) and 1(e). Each two bands centered around the $\Gamma$ point originating from B-C $\sigma$ states form a regularly small circular sheet and a quasi-hexagon sheet, termed $\sigma_{1}$ and $\sigma_{2}$ sheets, respectively. Another large quasi-hexagon sheet dominated by B-C $\pi$ state arises from the band crossing $M$ and $K$ points. The band, centered around $K$ point, originates from hybridized states of B-C $\pi$ and K $d$ orbitals, forming six small $\mu$ sheets. The presence of obviously distinct FS indicates that the investigated systems may be candidates with separate superconducting gaps. 

\begin{table}[ptb!]
	\caption{Parity eigenvalues of wave function of 11 occupied bands at four TRIMs and the $\mathbb{Z}_{2}$ index of AB-KB$_{2}$C$_{2}$.}
	\scalebox{1.06}{
		\begin{tabular}{cccccccccccccccccccccccccc}
			\hline
			\hline
			TRIMs&\multicolumn{11}{c}{Parities of bands}&Products&$\mathbb{Z}_{2}$ \\
			\hline
			$\Gamma$&+1&-1&-1&-1&+1&-1&+1&-1&+1&+1&-1 &+1&    \\
			$M_{0}$ &-1&+1&+1&+1&-1&+1&-1&+1&-1&-1&+1 &-1&1    \\
			$M_{1}$ &-1&+1&+1&+1&-1&+1&-1&+1&-1&-1&+1 &-1&    \\
			$M_{2}$ &-1&+1&+1&+1&-1&+1&-1&+1&-1&-1&+1 &-1&    \\
			\hline
		\end{tabular}
	}
\end{table}

In contrast to the band structure of AA-KB$_{2}$C$_{2}$, a Dirac point appears at the $K$ point in that of AB-KB$_{2}$C$_{2}$, which is determined by the stacking arrangement of B-C layers. The inclusion of spin-orbit coupling (SOC) results in the opening of a small gap of about 3 meV at the $K$ point, as depicted in Fig. S6(g). In addition, the entire continuous band gap under the Dirac point emerges, as shown in Figs. S6(e) $\sim$ S6(j). Consequently, we observe topological nature of AB-KB$_{2}$C$_{2}$ with a $\mathbb{Z}_{2}$ index of `1' in $k_{3}$ = 0.0. Furthermore, to further confirm the topological index $\mathbb{Z}_{2}$, we analyze the parity eigenvalues of the wave function at four time-reversal momentum points (TRIMs) of eleven bands. For 2D hexagonal lattice, these points include $\Gamma$ (0.0, 0.0), and other three additional equivalent points $M_{0}$ (0.5, 0.0), $M_{1}$ (0.0, 0.5) and $M_{2}$ (0.5, 0.5) \cite{PhysRevB.107.235154}. The $\mathbb{Z}_{2}$ index can then be calculated using the formula: (-1)$^{\mathbb{Z}_{2}}$ = $\prod_{1}^{4}$$\prod_{i=1}^{N_{occ.}}$$\delta(\Gamma_{i}$), where $\delta(\Gamma_{i})$ represents the parity eigenvalues of the occupied bands below the gap at TRIMs. The calculated data in Table I again confirm the electronic topological nontrivial nature. The high-symmetry points of edge states are recompiled with $\bar{\Gamma}$, $\bar{M}$, and $\bar{K}$, as illustrated in Fig. S6(a). Furthermore, the edge states on the (100) surface are shown in Figs. 2(a), S6(c), and S6(d). It is evident that the edge states above the $E_{F}$ 0.6 eV start and split from the small gap at the $K$ point, and then, melt and merge into the bulk bands along the $\bar{K}$-$\bar{M}$ direction. As shown in Fig. 2(b), the $k_{x}$-$k_{y}$ plane crossing the edge states presents six trigonal sections around $K$ points, as coordinated with the edge states in Fig. 2(a). In experiment, one can raise the $E_{F}$ by injecting of electrons or applying pressures to further detect such topological nature with angle resolved photoemission spectroscopy (ARPES) measurements \cite{PhysRevLett.127.266401,Li2021,Kang2019,PhysRevB.106.214527}.  

The spin texture of the highest occupied band throughout the whole BZ is shown in Fig. 2(c). It is known that under the Rashba-type spin splitting, spins are polarized within the plane of sample, while with the Ising-type spin splitting, spins become polarized out of the plane of sample \cite{PhysRevB.109.165424}. Consequently, the spin splitting of the band at $K^{'}$ [Fig. S6(k)] and $K$ [Fig. S6(l)] points corresponds to the Ising type, while at $\Gamma$ point [Fig. S6(m)], it manifests as a combination of Rashba and Ising types. The Ising-type spin splitting opposing in the $K$ and $K^{'}$ valleys is referred as Zeeman spin splitting \cite{PhysRevB.109.165424}. The coexistence of different types of spin splitting is rare \cite{PhysRevB.109.165424} and warrant further investigation. Moreover, the in-plane spin component forms hexagonal and trigonal vortices around $\Gamma$ and $K$ ($K^{'}$) points, respectively, while the antivortices appear around the $M$ point. When only considering the in-plane spin component, the BZ resembles a torus and is considered as a tangential field \cite{PhysRevLett.128.166601}. According to the Poincare-Hopf (PH) theorem \cite{Poincare-Hopf}, the Euler number of a torus, which is zero, equals the sum of winding numbers around spin vortices. At the points $\Gamma$, $K$, and $K^{'}$, there are three spin vortices, each with a winding number of +1. Conversely, the winding numbers of antivortices around three $M$ points are all -1. The total winding number in the whole BZ cancels out to zero. Therefore, as per the PH theorem \cite{Poincare-Hopf}, the presence of spin antivortex is crucial, which also points to the topological nature \cite{PhysRevLett.128.166601}.

$Phononic$ $properties.$ It is important to analyze lattice vibration using atomic-vibrational-mode-resolved phonon spectra and the phonon density of states (PHDOS), as illustrated in Fig. S7. In both AA- and AB-KB$_{2}$C$_{2}$ structures, the distributions of atomic vibration are remarkably similar. The whole frequency range extends from 0 to about 123 meV, where the upper limit is determined by the vibrations of B/C atoms with smaller masses. Consequently, the vibrations predominantly occur at higher frequencies above 25 meV and are notably prominent above 42 meV. The out-of-plane vibrations of B/C atoms range from $\sim$30 to $\sim$90 meV, while the in-plane cover the whole frequency. In contrast, the vibrations of K atoms with heavier masses predominantly occur at lower frequencies below 42 meV. The first and second/third acoustic branches originate from the out-of-plane and in-plane vibrations, respectively. The distributions of atomic vibration are distinct, and no distinguished phonon gap exists. Additionally, several slightly soft modes, marked by light blue ellipses, appear and are dominated by B/C in-plane vibration modes. 

Based on the calculated tight-binding parameters \cite{TOGO20151,PhysRevLett.120.016401} and the surface Green's function, the phononic local density of states (LDOS) \cite{Sancho_1984,Sancho_1985,PhysRevLett.120.016401,WU2017} are obtained. As shown in Fig. 2(d), we observe several edge LDOS forming around $\bar{\Gamma}$ and $\bar{M}$ points. There are two LDOS about 66 meV splitting from the $\bar{\Gamma}$ point and merging into bulk bands. Around $\bar{M}$ point, two crisscrossing long LDOS extends from $\sim$54.5 to $\sim$66.5 meV, with an isolated short LDOS ranging from $\sim$66.5 to $\sim$66.7 meV. Figures 2(e) and 2(f) display the corresponding isofrequency surfaces contours, marked by red dotted lines on the (001) plane at 64 and 65.5 meV. The above analyses suggest that the investigated system exhibits phononic topological behaviors, which are advantageous for phonon transport \cite{PhysRevB.81.064301} and interfacial superconductivity \cite{PhysRevB.94.081116,PhysRevB.97.085142}. The predicted phononic topology \cite{PhysRevLett.120.016401} may be demonstrated by various techniques \cite{Bruesch,PhysRevB.91.094307}, such as electron energy loss spectroscopy \cite{10.1063/1.4928215}. 

\begin{figure*}[hptb]
	\centering
	\includegraphics[width=1\linewidth]{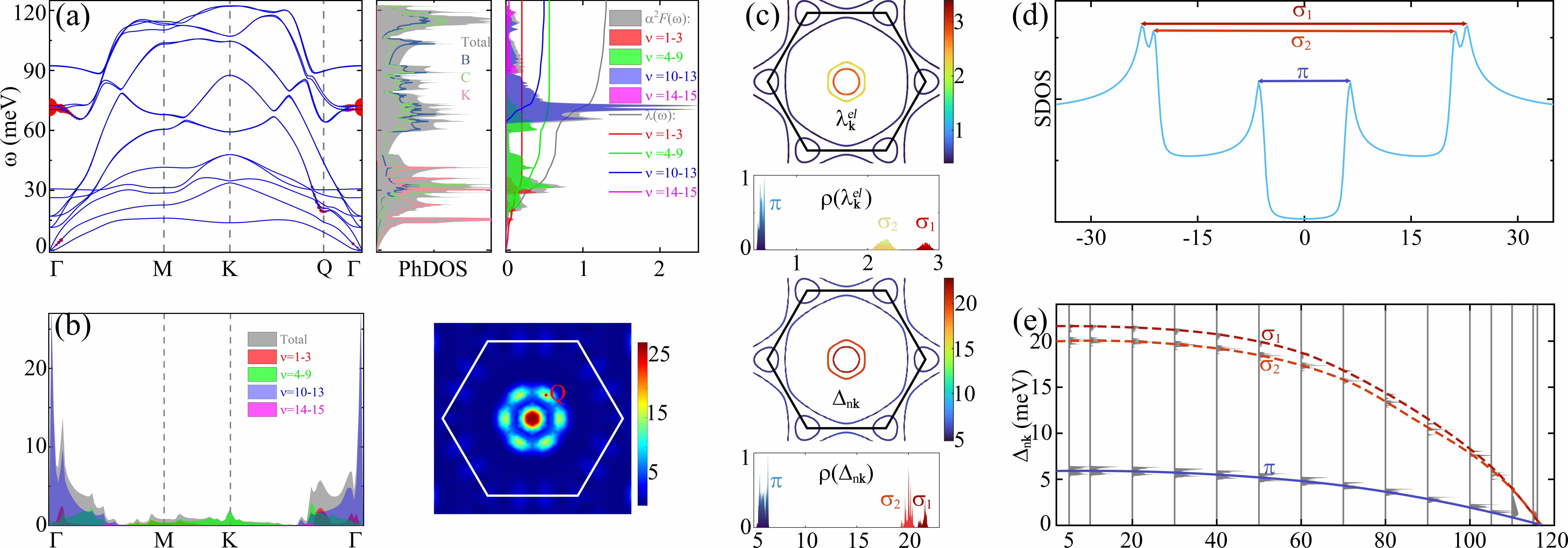}
	\caption{The related superconducting properties for AA-KB$_{2}$C$_{2}$. (a) Phonon weighted with the EPC parameters $\lambda^{\mathit{ph}}_{\mathbf{q}}$ (red shadows), PhDOS, and the total cumulative EPC $\lambda(\omega)$, calculated by formula $\lambda(\omega)$=2$\int\alpha^{2}\mathit{F}(\omega)/\omega d\omega$, with the mode-resolved Eliashberg spectral function $\alpha^{2}\mathit{F}(\omega)$. (b) The phononic EPC parameters $\lambda^{\mathit{ph}}_{\mathbf{q}}$ along the high-symmetry route $\Gamma$-$M$-$K$-$\Gamma$ and projected on the BZ. (c) The electronic EPC parameters $\lambda^{\mathit{el}}_{\mathbf{k}}$ on the FS with the strength distribution $\rho(\lambda^{\mathit{el}}_{\mathbf{k}})$ and momentum-resolved superconducting gap $\Delta_{n\mathbf{k}}(\omega=0)$ on the FS with the strength distribution of gap $\rho(\Delta_{n\mathbf{k}})$ at $T$ = 5 K. (d) SDOS at $T$ = 5 K. (e) Energy distribution of the gaps $\Delta_{n\mathbf{k}}$ versus $T$.}
	\label{fig:fig}
\end{figure*}

$EPC$ $and$ $Superconductivity.$ The superconducting behaviors in AA- and AB-KB$_{2}$C$_{2}$ are presented in Figs. 3 and S9. Remarkably, the large values of $\lambda^{\mathit{ph}}_{\mathbf{q}}$ are located at the $\Gamma$ point around 71 meV in the $\nu=10\sim13$ optical branches with $E$ vibration modes (Table S5), dominated by the in-plane vibrations of B/C atoms (Figs. S7(a), S7(b), S8, and Table S5). In addition, the acoustic phonon vibration modes with wave vector $\mathbf{q}$ $\approx$ 0.3 $\times$ $\Gamma K$, namely Q point, suffer slight softening, leading to a small value of $\lambda^{\mathit{ph}}_{\mathbf{q}}$ on the phonon dispersions and an obvious peak value of EPC $\lambda^{\mathit{ph}}_{\mathbf{q}}$ in the BZ [Figs. 3(b) and S9(b)]. These special points, where the $\lambda^{\mathit{ph}}_{\mathbf{q}}$ congregates, usually trigger high peaks of Eliashberg functions $\alpha^{2}\mathit{F}(\omega)$ and contribute to the accumulation of EPC constant $\lambda(\omega)$. Particularly, the optical $E$ modes with in-plane B-C vibrations give rise to the highest peak of $\alpha^{2}\mathit{F}(\omega)$ and a jumping increase of $\lambda(\omega)$. These results are also supported by the mode-resolved $\alpha^{2}\mathit{F}(\omega)$ and $\lambda(\omega)$, which exhibit that the three acoustic branches ($\nu=1\sim3$), the four branches with the largest value of $\lambda^{\mathit{ph}}_{\mathbf{q}}$ ($\nu=10\sim13$), and the two high-energy optical branches ($\nu= 14\sim15$) contribute 15.84(14.47) \%, 37.82(39.63) \%, and 3.03(3.13) \% of the total $\lambda(\omega)$ for AA(AB)-KB$_{2}$C$_{2}$, respectively. The contribution of the remaining modes ($\nu =4\sim9$) following the vibration of PHDOS is up to 43.31(42.77) \%. Each branch in $\nu=1\sim3, 4\sim9, 10\sim13,$ and $14\sim15$ contributes 5.3(4.8) \%, 7.2(7.1) \%, 9.5(9.9) \%, and 1.5(1.6) \% of the total $\lambda(\omega)$, respectively, indicating again that the optical $E$ vibration modes with in-plane B-C vibrations play essential roles in superconductivity. Finally, the total EPC constants $\lambda(\omega)$ are 1.30 and 1.26 for AA- and AB-KB$_{2}$C$_{2}$, respectively.

With insight into the nature of superconductivity, we further investigate the FS with EPC parameters $\lambda^{\mathit{el}}_{\mathbf{k}}$ and superconducting gap, superconducting DOS (SDOS), and the temperature-dependent energy distribution of the gaps $\Delta_{n\mathbf{k}}$, calculated from the fully anisotropic ME theory \cite{migdal1958interaction,eliashberg1960interactions,Philip,PhysRevB.87.024505}. The EPC parameters $\lambda^{\mathit{el}}_{\mathbf{k}}$ are related to, but different from, $\lambda^{\mathit{ph}}_{\mathbf{q}}$ \cite{Jesse,PhysRevLett.119.087003,PhysRevB.101.104507}. Figures 3(c) and S9(c) show the EPC parameters $\lambda^{\mathit{el}}_{\mathbf{k}}$ dominate on the $\sigma_{1}$ and $\sigma_{2}$ sheets with strength distributions $\rho (\lambda^{\mathit{el},\sigma_{1}}_{\mathbf{k}})$ from 2.65 (2.58) to 2.94 (2.84) and $\rho(\lambda^{\mathit{el},\sigma_{2}}_{\mathbf{k}})$ from 2.01 (1.95) to 2.43 (2.32) for AA(AB)-KB$_{2}$C$_{2}$, respectively. Conversely, smaller values of $\lambda^{\mathit{el}}_{\mathbf{k}}$ are observed on the $\pi$ sheet \cite{note,PhysRevB.104.174519}, with corresponding strength distributions $\rho (\lambda^{\mathit{el},\pi}_{\mathbf{k}})$ ranging from 0.44 (0.42) to 0.56 (0.59). Overall, the EPC parameters $\lambda^{\mathit{el}}_{\mathbf{k}}$ range widely from 0.44 (0.42) to 2.94 (2.84). In Figs. 3(b) and S9(b), the separation of strength distributions on the $\sigma_{1}$ and $\sigma_{2}$ sheets arises from the extra hexagon-like distribution of $\lambda^{\mathit{ph}}_{\mathbf{q}}$ on $\sigma_{1}$ sheet \cite{PhysRevB.101.104507}. At $T$= 5 K, the regions where larger superconducting gaps open on FS coincide with those characterized by stronger EPC parameters $\lambda^{\mathit{el}}_{\mathbf{k}}$. This means that the strongest gap $\Delta^{\sigma_{1}}_{n\mathbf{k}}$, the secondary gap $\Delta^{\sigma_{2}}_{n\mathbf{k}}$, and the weakest gap $\Delta^{\pi}_{n\mathbf{k}}$ predominantly introduced by $\pi$ states of B/C atoms open on the $\sigma_{1}$, $\sigma_{2}$, and $\pi$ sheets, respectively. The ranges of three gaps $\Delta^{\sigma_{1}}_{n\mathbf{k}}$, $\Delta^{\sigma_{2}}_{n\mathbf{k}}$, and $\Delta^{\pi}_{n\mathbf{k}}$ at $T$= 5 K are 20.91 (20.37) $\sim$ 22.23 (21.47), 19.23 (18.65) $\sim$ 20.76 (20.18), and 5.00 (4.96) $\sim$ 6.48 (7.06) meV, respectively. Hence, the gap divisions arising from the $\Delta^{\sigma_{1}}_{n\mathbf{k}}$, $\Delta^{\sigma_{2}}_{n\mathbf{k}}$ and $\Delta^{\sigma_{2}}_{n\mathbf{k}}$, $\Delta^{\pi}_{n\mathbf{k}}$ are 0.15 (0.19) and 12.75 (11.59) meV, respectively. Combining the electronic orbitals on the FS in Fig. 1 with the distribution results of $\lambda^{\mathit{el}}_{\mathbf{k}}$ and $\Delta_{n\mathbf{k}}$, one thing to emphasize that $\sigma$-states electron of B/C atoms plays a significant role in EPC. To summarize, superconductivity in AA- and AB-KB$_{2}$C$_{2}$ stems from EPC between $\sigma$-states electrons and in-plane phonon vibrations of B/C atoms. 

Furthermore, scanning tunneling spectroscopy measurements can correctly reveal prominent features in the SDOS below $T_{c}$, yielding persuasive evidence of multi-gap superconductivity \cite{PhysRevLett.119.087003,PhysRevLett.101.166407,PhysRevB.96.094510,MIavarone_2003}. The SDOS is calculated using the formula $\dfrac{N_{S}(\omega)}{N_{F}}=Re[\omega/\sqrt{\omega^{2}-\Delta^{2}(\omega)}]$, where `Re' represents the gap $\Delta(\omega)$ projected on the real axis, $N_{S}(\omega)$ and $N_{F}$ are the electron DOS of superconducting and normal states at $E_{F}$, respectively. As depicted in Figs. 3(d) and S9(d), the SDOS at $T$ = 5 K exhibit three distinct peaks corresponding to three gaps $\Delta^{\sigma_{1}}_{n\mathbf{k}}$, $\Delta^{\sigma_{2}}_{n\mathbf{k}}$, and $\Delta^{\pi}_{n\mathbf{k}}$, providing convincing evidence of the existence of three-gap superconductivity. The results are consistent with the aforementioned analyses. Figures 3(e) and S9(e) illustrate the changes in the gaps $\Delta_{n\mathbf{k}}$ versus $T$, indicating a temperature dependence characteristic of the standard BCS type. Furthermore, at each temperature, the gaps $\Delta_{n\mathbf{k}}$ separate three distinct values, confirming again the nature of three-gap superconductivity. The $T_{c}$ is defined as the maximum temperature at which the $\Delta_{n\mathbf{k}}$ gradually vanish and converge. Therefore, the $T_{c}$ in the AA- and AB-KB$_{2}$C$_{2}$ reach up to 116.5 and 112.5 K, respectively, significantly exceeding those of monolayer Mg$_{2}$B$_{4}$C$_{2}$ (\textless 50 K) \cite{Singh2022}, monolayer MgB$_{2}$ under 4 \% strains (\textgreater 50 K) \cite{PhysRevB.96.094510}, four-gap superconductor MgB$_{4}$ film ($\sim$ 52 K) \cite{PhysRevB.101.104507}, monolayer LiBC (70 K) \cite{PhysRevB.104.054504}, three-gap superconductors LiBCH (\textgreater 80 K) \cite{liu2024threegapsuperconductivitytc80}, surface superconductors Ca$_{n}$B$_{n+1}$C$_{n+1}$ (n = 1, 2, 3, $\cdots$) ($\sim$ 90 K) \cite{doi:10.1021/acs.nanolett.2c05038}, LiB$_{2}$C$_{2}$ film (92 K) \cite{PhysRevB.101.094501}, hydrogenated monolayer MgB$_{2}$ under 5 \% strains ($\sim$ 103 K) \cite{PhysRevLett.124.077001}. It is noteworthy that both $T_{c}$ exceed 77 K, which is highly significant and clearly advantageous for potential further applications. The values of $T_{c}$ set new records among all currently known 2D ambient-pressure superconductors, which is a remarkable achievement in the field of superconductivity research.

In brief, the essential SOC in electronic topology requires heavier atoms, which results in smaller Debye frequencies that are not beneficial for high-$T_{c}$ superconductivity \cite{PhysRev.108.1175,PhysRevLett.124.077001,PhysRevLett.86.4656}. Consequently, the coexistence of nontrivial topology and superconductivity with such a high $T_{c}$ is reported for the first time, potentially offering more opportunities to observe topological superconductors \cite{PhysRevB.106.214527,Singh2022,PhysRevB.105.024517,npjcm-si,PhysRevB.107.235154} and Majorana zero modes \cite{PhysRevB.107.235154,Liu2020,Kong2021}.

$Effect$ $of$ $BTS$ $on$ $superconductivity.$ Due to significant enhancement of $T_{c}$ through strain engineering or lattice deformation in numerous 2D superconductors \cite{PhysRevLett.111.196802,PhysRevB.100.094516,PhysRevB.96.094510,PhysRevB.101.094501,PhysRevLett.124.077001}, we consider the effects of BTS on the predicted three-gap superconductors, described by $\varepsilon$=($a$-$a_{0}$)/$a_{0}$ $\times$ 100 \%, here, $a_{0}$ and $a$ are the lattice constants before and after BTS. We conduct calculations using BTS until the point where conspicuous imaginary frequencies appear. According to the results in Fig. S10, AA- and AB-KB$_{2}$C$_{2}$ remain dynamically stable until the strains exceed $\varepsilon$ = 3.5 \% and 3.4 \%, respectively. With increasing BTS, as illustrated in Figs. S11(a) and S11(f), both the overall frequency and the strongly coupled phonon modes experience a red shift and sharp softening, especially $E$ modes shift from $\sim$71 to $\sim$38 meV, and there is noticeable softening of the acoustic modes around Q point. Furthermore, Figs. S11(b) and S11(g) also exhibit that the large values of $\lambda^{\mathit{ph}}_{\mathbf{q}}$ evidently appear at the $E$ modes around $\Gamma$ point and the acoustic modes around Q point, where the softening modes are still predominantly influenced by the in-plane vibrations of B/C atoms (Figs. S7 and S8). Like other similar graphite-like structures \cite{PhysRevB.101.104507,PhysRevB.104.174519,PhysRevB.101.094501,PhysRevB.100.094516,liu2024threegapsuperconductivitytc80}, the anomalous large values of $\lambda^{\mathit{ph}}_{\mathbf{q}}$ around softening-modes Q point derive from Kohn anomalies \cite{PhysRevLett.2.393} rather than Fermi surface nesting effect (Fig. S12). The results show increased proportions of $\nu$=1$\sim$3 in the total $\lambda$ [Figs. S11(a) and S11(f), indicating the significant role of low-frequency phonon in BTS-cases EPC. Consequently, compared to the strain-free cases, the EPC constants $\lambda$ for AA- and AB-KB$_{2}$C$_{2}$ are boosted up to 2.86 and 3.09, respectively. Moreover, as depicted in Fig. S13, the BTS does not dramatically impact the band structures and enhance the electronic DOS at $E_{F}$, indicating that there are no additional electrons contributing to EPC under BTS cases. Thus, in the increase of EPC, phonon plays a more beneficial role in BTS-cases EPC and in the formation of electron Cooper pairs, rather than electrons \cite{PhysRevLett.111.196802,PhysRevB.100.094516,PhysRevB.96.094510,PhysRevB.101.094501,PhysRevLett.124.077001}. This evidently suggests that phonon is significant media in BCS superconductors. Moreover, the regions with large values of $\lambda^{\mathit{el}}_{\mathbf{k}}$ ($\Delta_{n\mathbf{k}}$), and the robust three-gap superconducting nature are consistent with the strain-free cases, as shown in Figs. S11(c), S11(d), S11(h), and S11(i). In Figs. S11(e) and S11(j), the corresponding critical temperature as high as $T_{c}$ = 153 and 157 K, which exceed the half of 300 K. By using a substrate with an adjustable lattice constant \cite{PhysRevB.73.024509} or a suspended monolayer \cite{PhysRevB.96.094510,PhysRevLett.111.196802,C5NR07755A}, BTS can potentially achieve larger values of $\Delta_{n\mathbf{k}}$ and elevate $T_{c}$ beyond 153 K.

$Conclusions.$ In conclusion, two new sandwich configurations of KB$_{2}$C$_{2}$ with dynamical stability are identified from various B-C layers constructions. Their stabilities, electronic/phononic, and superconducting properties are throughly investigated. Electrons at $E_{F}$ form several obviously distinct FS corresponding to different atomic orbitals, potentially leading to multi-gap superconductivity. Moreover, the topological behaviors of electrons and phonons appear. Meanwhile, Ising-type spin splitting emerges in AB-KB$_{2}$C$_{2}$. Using anisotropic ME equations, the results suggest that phononic EPC parameters $\lambda^{\mathit{ph}}_{\mathbf{q}}$ primarily localize around high-frequency optical $E$ modes near the $\Gamma$ point, with a minor contribution from acoustic modes showing slight softening around the Q point. With the strong and evident three-region distribution characteristic of electronic EPC parameters $\lambda^{\mathit{el}}_{\mathbf{k}}$, an exciting robust three-gap superconducting nature with high $T_{c}$ above 112 K is discovered. Furthermore, the EPC arises from the coupling between optical $E$ modes due to in-plane vibration of B/C and $\sigma$-states electrons originating from $p_{x,y}$ orbitals of B/C. Moreover, under the BTS, $T_{c}$ can be further increased to above 153 K. Distinctively, phonon plays a more important role in the BTS-cases EPC, rather than electrons. To sum up, our research offers intriguing sandwich structures aimed at discovering higher-$T_{c}$ 2D superconductors, while also holding promise for exploring the fundamental quantum physics.

\section*{ACKNOWLEDGEMENTS}
Thanks to Xiao-Yan Cui for assisting with the calculated submissions and to the computing center (old area) of the Institute of Applied Physics and Computational Mathematics for providing computational resources. This work was supported by the National Natural Science Foundation of China (Grant Nos. 12175023, 12074213, 11574108), the Major Basic Program of Natural Science Foundation of Shandong Province (Grant No. ZR2021ZD01), and the Project of Introduction and Cultivation for Young Innovative Talents in Colleges and Universities of Shandong Province.

\bibliography{KB2C2.bib}

\begin{thebibliography}{78}
\bibitem [{not()}]{note}%
  \BibitemOpen
  \href@noop {} {\bibinfo  {journal} {Actually, there are not only $\pi$ sheets
  but also $\mu$ sheets. However, the electrons on the hybridized-inclued $\mu$
  sheets are too weak (Figs. 1(b) and 1(e)) to open a new superconducting gap
  (Figs. 3(d) and 3(i)). Therefore, the smallest superconducting gaps $\Delta$
  are opened by $\pi$ electrons and named with $\Delta^{\pi}$ in the paper.}\
  }\BibitemShut {NoStop}%
\end{thebibliography}
\end{document}